\documentclass[prl,aps, reprint,superscriptaddress, amsmath,amssymb,floatfix, nofootinbib]{revtex4-2}

\usepackage{graphicx}
\usepackage{dcolumn}
\usepackage{bm}
\usepackage{float}
\usepackage{siunitx}
\usepackage{braket}
\usepackage[dvipsnames]{xcolor}
\usepackage[colorlinks=true,linkcolor=red,citecolor=blue]{hyperref}

\begin{document}

\title{Native qudit entanglement in a trapped ion quantum processor}

\author{Pavel Hrmo }
\thanks{These authors contributed equally to this work.}
\affiliation{Institut f\"ur Experimentalphysik, Universit\"at Innsbruck, Technikerstraße 25/4, 6020 Innsbruck, Austria}
\author{Benjamin Wilhelm }
\thanks{These authors contributed equally to this work.}
\affiliation{Institut f\"ur Experimentalphysik, Universit\"at Innsbruck, Technikerstraße 25/4, 6020 Innsbruck, Austria}
\author{Lukas Gerster}
\affiliation{Institut f\"ur Experimentalphysik, Universit\"at Innsbruck, Technikerstraße 25/4, 6020 Innsbruck, Austria}
\author{Martin W. van Mourik}
\affiliation{Institut f\"ur Experimentalphysik, Universit\"at Innsbruck, Technikerstraße 25/4, 6020 Innsbruck, Austria}
\author{Marcus Huber}
\affiliation{Atominstitut, Technische Universit{\"a}t Wien, 1020 Vienna, Austria}
\affiliation{Institute for Quantum Optics and Quantum Information - IQOQI Vienna, Austrian Academy of Sciences, Boltzmanngasse 3, 1090 Vienna, Austria}
\author{Rainer Blatt}
\affiliation{Institut f\"ur Experimentalphysik, Universit\"at Innsbruck, Technikerstraße 25/4, 6020 Innsbruck, Austria}
\affiliation{Institut für Quantenoptik und Quanteninformation,
Österreichische Akademie der Wissenschaften, Technikerstraße 21a, 6020 Innsbruck, Austria}
\affiliation{AQT, Technikerstraße 17, 6020 Innsbruck, Austria}
\author{Philipp Schindler}
\affiliation{Institut f\"ur Experimentalphysik, Universit\"at Innsbruck, Technikerstraße 25/4, 6020 Innsbruck, Austria}
\author{Thomas Monz}
\affiliation{Institut f\"ur Experimentalphysik, Universit\"at Innsbruck, Technikerstraße 25/4, 6020 Innsbruck, Austria}
\affiliation{AQT, Technikerstraße 17, 6020 Innsbruck, Austria}
\author{Martin Ringbauer}
\affiliation{Institut f\"ur Experimentalphysik, Universit\"at Innsbruck, Technikerstraße 25/4, 6020 Innsbruck, Austria}

\begin{abstract}
    Quantum information carriers, just like most physical systems, naturally occupy high-dimensional Hilbert spaces. Instead of restricting them to a two-level subspace, these high-dimensional (qudit) quantum systems are emerging as a powerful resource for the next generation of quantum processors. Yet harnessing the potential of these systems requires efficient ways of generating the desired interaction between them.
    Here, we experimentally demonstrate an implementation of a native two-qudit entangling gate in a trapped-ion qudit system up to dimension $5$. This is achieved by generalizing a recently proposed light-shift gate mechanism to generate genuine qudit entanglement in a single application of the gate. The gate seamlessly adapts to the local dimension of the system with a calibration overhead that is independent of the dimension.
\end{abstract}

\maketitle

\begin{figure}[t]
    \begin{center}
    \includegraphics[width=0.85\columnwidth]{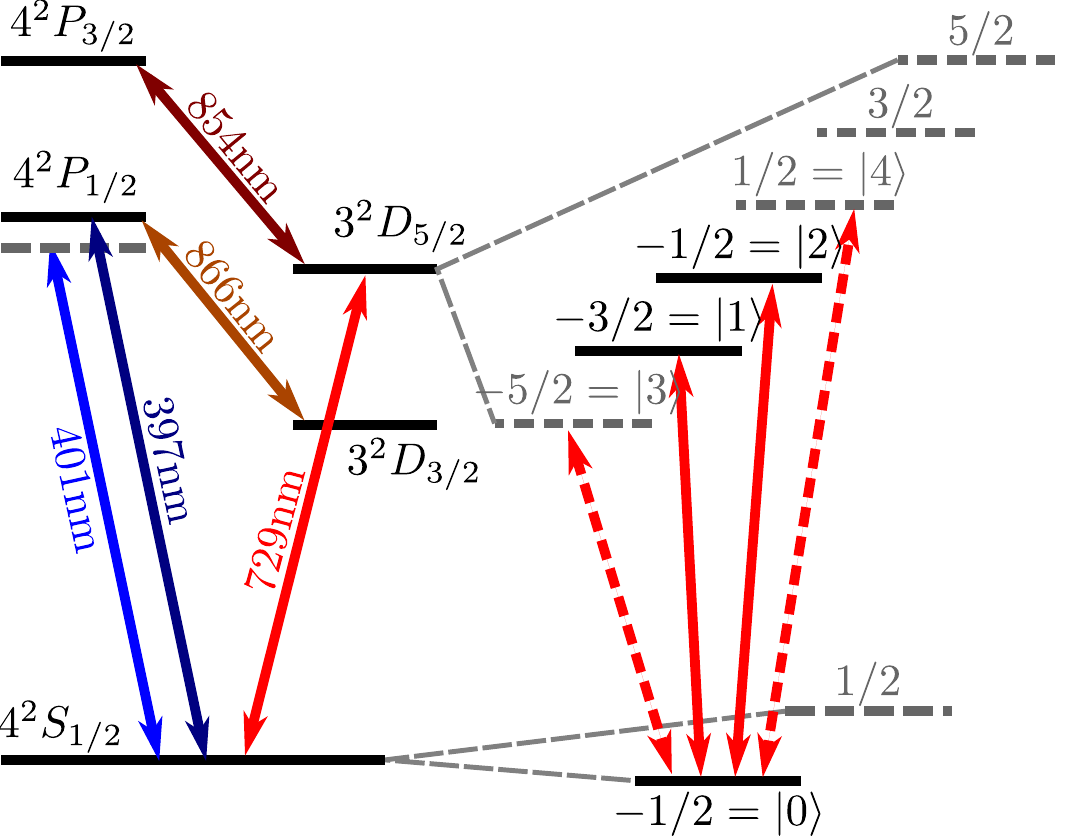}
    \caption{\textbf{Level scheme of the $^{40}$Ca$^+$ ion.} Encoding quantum information in the sub-levels of the $S_{1/2}$ and $D_{5/2}$ manifold allows us to increase the size of the computational Hilbert space. Coherent operations between the sub-levels can be performed using 729 nm laser light, while  401 nm light is used to generate the light shift for the state dependent force.}
    \label{fig:levelscheme}
    \end{center}
\end{figure}

Quantum computing has taken great strides in the past decades with multiple platforms demonstrating control over tens of qubits \cite{Saffman2019,Huang2020,Bruzewicz2019a,Slussarenko2019}. However, scaling these systems to a regime beyond the capabilities of classical computers remains extremely challenging, both in terms of increasing the size of the quantum computational Hilbert space, and in terms of increasing the depth of the computational circuits. 
Importantly, addressing these challenges does not necessarily require increasing the number of quantum particles, or reducing gate errors. Instead, significant potential for progress lies in plain sight when appreciating that our quantum information carriers are multi-level, not two-level systems. 
A natural way to extend the computational Hilbert space and the circuit depth, without increasing the complexity of quantum devices is thus to use the full multi-level or \emph{qudit} structure of existing quantum information carriers such as trapped ions, see Fig.~\ref{fig:levelscheme}. 

Qudit control has already been demonstrated in a number of architectures~\cite{Chi2022,Ringbauer2017Coherence,Wang2018,Ahn2000,Godfrin2017,Anderson2015,Gedik2015,Kononenko2020,Morvan2020,Zhang2013Contextuality,Hu2018a,Blok2021}, including trapped ions~\cite{Ringbauer2021,Senko2015,Leupold2018,Malinowski2018}. Qudit approaches not only enable reduced circuit complexity~\cite{Lanyon2008} and simplifications of virtually any quantum circuit~\cite{Wang2020}, but also benefit from more powerful quantum error correction~\cite{Campbell2012,Duclos-Cianci2013,Anwar2014,Watson2015}, and allow for native quantum simulation of natural systems such as lattice gauge models~\cite{Banuls2020} and quantum chemistry~\cite{MacDonell2020}. The key to these advantages, however, is an appropriate set of quantum operations. While for qubits all entangling gates are equal up to local rotations, the same is not true for qudits, where the richer Hilbert space allows for different forms of coherence~\cite{Ringbauer2017Coherence} and entanglement~\cite{Kraft2018}. We can roughly classify qudit entangling gates by their \emph{entangling power}, i.e.\ the amount of entanglement that is created relative to a maximally entangled state of two qubits. While in principle already the simplest gate, a qubit entangling gate embedded in a higher-dimensional Hilbert space, would suffice for universal qudit quantum computation~\cite{Brennen2006,Ringbauer2021}, having access to a range of gates with different entangling power will be crucial for unlocking the full potential of a qudit quantum processor. Not only will a diverse set of interactions enable the direct simulation of a wider range of physical systems, but it will also enable much more efficient quantum circuit decomposition. 

Here we describe and demonstrate a native qudit entangling gate in a trapped-ion quantum processor. Being based on differential light shifts between the ground- and excited state manifolds on an optical transition, the gate action can be made symmetric on all excited qudit states. We show that this implies that the same gate mechanism can be used irrespective of the local dimension of the quantum system. Crucially, this means that the experimental calibration overhead does not increase with qudit dimension. We further show that direct application of the  gate can generate maximal qudit entanglement up to dimension $4$, and we outline generalizations to shape the gate action for maximal entanglement in higher dimensions. We characterize the gate dynamics and noise sources in detail, demonstrating that this gate can be a stepping stone into the world of native qudit quantum information processing with trapped ions.

\begin{figure}[htb]
    \begin{center}
    \includegraphics[width=0.9\columnwidth]{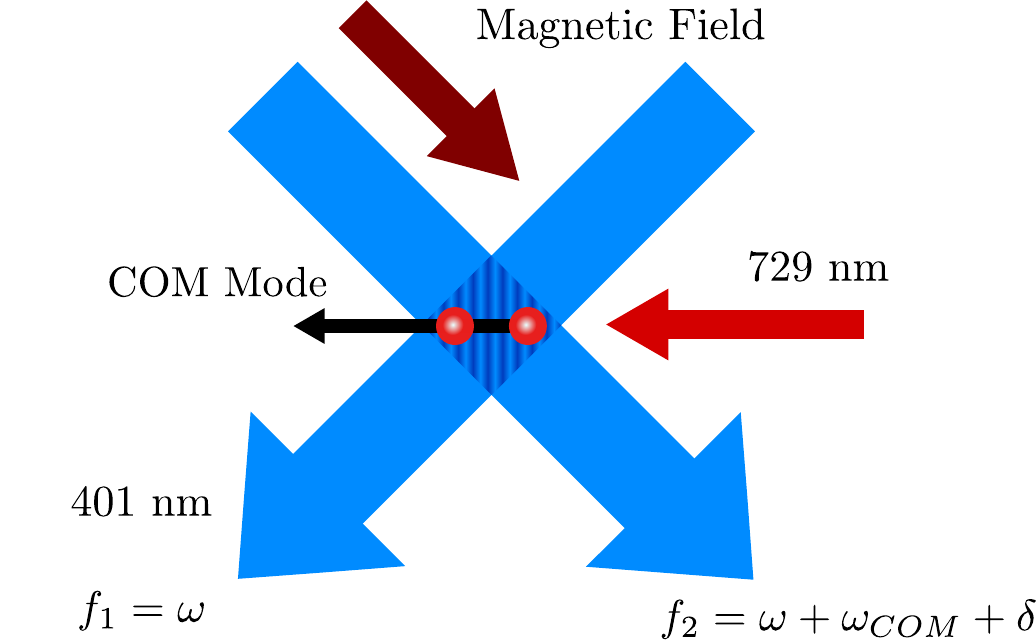}
    \caption{\textbf{Experimental setup.} Two beams with $\lambda = \SI{401}{nm}$ and frequencies $f_1, f_2$ with a relative detuning of $\omega_\mathrm{COM}+\delta$ are intersected at a $90^\circ$ angle to drive the spin-dependent force. A $\SI{729}{nm}$ laser along the axial trap direction is used to implement local operations.}
    \label{fig:beamgeometry}
    \end{center}
\end{figure}

The principle behind the qudit entanglement generation is the application of light-shift (LS) gates, in which a state-dependent optical-dipole force couples the ions' spin to their common motion in the trap. Light-shift gates have been well studied for entangling hyperfine and Zeeman qubits with a pair of intersecting laser beams (see Fig.~\ref{fig:beamgeometry}) that create a travelling wave~\cite{Leibfried2003, Ballance2016,Hughes2020,Baldwin2021}. 
The travelling wave produces a spatially modulated light shift that drives an excursion in the phase space of one of the motional modes.  The ions acquire a different geometric phase depending on their spin state, leaving them in an entangled state after completing the excursion. The beams also introduce additional, unwanted local phases as the spin states experience different light shifts.
For hyperfine or Zeeman qubits, the polarisation of the optical beams can be used to null these differential light shifts on the spin states. 
In practice, however, these gates are typically applied with a \emph{spin-echo} pulse, inserted between two halves of the LS gate pulse during which the ions complete a single loop in motional phase space. 
This ensures that any residual light shifts between the two spin states due to experimental imperfections are cancelled. Adding these local $\pi$ pulses has the further advantage of suppressing slow qubit frequency drifts and decoupling from the optical phase of the LS gate laser. 
Using this technique, the LS gate can also be implemented without the requirement to intrinsically null the differential light shift.
This opens the opportunity to apply the gate to qubits with an energy difference in the optical domain, without careful selection of the laser frequency. 
Such gates were theoretically described~\cite{Sawyer2021} and implemented~\cite{Clark2021} for an optical qubit formed by sub-levels of the $S_{1/2}$ and $D_{5/2}$ manifold in $^{40}$Ca$^+$ ions. 

We now describe how this gate scheme can be generalized to generate genuine qudit entanglement for qudits of arbitrary dimension encoded in the Zeeman sub-levels of the $S_{1/2}$ and $D_{5/2}$ manifolds of $^{40}$Ca$^+$ ions, see Fig.~\ref{fig:levelscheme}. 
Formally, the two-ion LS Hamiltonian, after adiabatic elimination of the excited states and application of the Lamb-Dicke and rotating-wave approximations, can be written in the interaction frame of the ions' center-of-mass (COM) motion as
\begin{equation}
    H_\mathrm{LS} = \dfrac{i\hbar\eta}{2}\sum_j\sum_N \Delta_{N,j} \ket{j_N}\bra{j_N}e^{-i\delta t}e^{i\varphi_N}a^\dag + \mathrm{h.c.} .
    \label{eq:LightShiftHamiltonian}
\end{equation}
Here $\Delta_{N,j}$ represents the light shift on the state $\ket{j}$ of ion $N$, $\delta$ the detuning from the motional mode frequency, $\eta$ the Lamb-Dicke parameter, $a^\dag$ the creation operator of the mode and $\varphi_N$ the phase determined by the inter-ion distance within the travelling wave of the laser. Integrating Eq.~\eqref{eq:LightShiftHamiltonian}, we obtain the propagator $U_\mathrm{LS}(t)$ describing a state-dependent force on the ions' motion that can be visualised as loops in motional phase space which will periodically return to the origin after a time $t_\mathrm{g} = 2\pi/\delta$. During the evolution in phase space, each initial spin state combination $\ket{jk}$ of the two ions will, in the most general case, pick up a different geometric phase $\phi_{jk}$ after a single LS gate pulse application.

\begin{figure*}[t]
    \begin{center}
    \includegraphics[width=\textwidth]{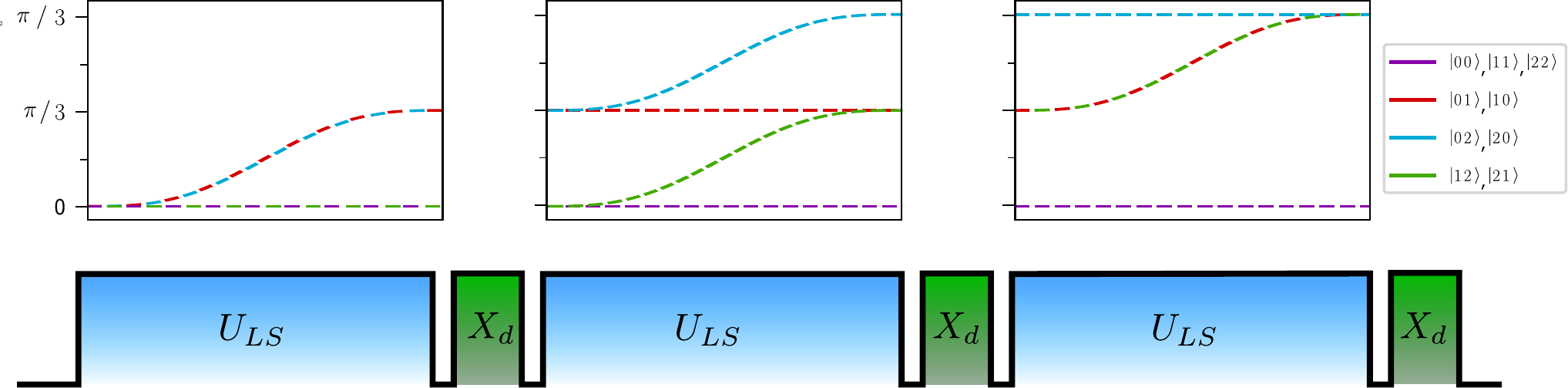}
    \end{center}
    \caption{\textbf{Top} Phase evolution of the two-qutrit state components relative to the $\ket{00}$ ground state during the application of the LS gate pulses. States $\ket{01}$ and $\ket{10}$ are shown in red, $\ket{02}$ and $\ket{20}$ in blue, and $\ket{21}$ and $\ket{12}$ in green. The equal spin states $\ket{00}, \ket{11}, \ket{22}$ in purple do not acquire any relative phases. \textbf{Bottom} Corresponding pulse scheme to implement $G(\theta)$ for a qutrit $(d=3)$. Light-shift gate pulses $U_\mathrm{LS}(t_\mathrm{g})$ are interlaced with cyclic permutation gates $X_3$.}
    \label{fig:gatescheme}
\end{figure*}

Similar to the qubit case, we can now symmetrize the geometric phases for qudits, while cancelling differential light shifts. We achieve this by  encoding our Qudits in the $\ket{0} = S_{1/2, m_j = -1/2}$ ground state and the Zeeman sub-levels of the $D_{5/2}$ manifold as $\ket{i}$, with $i\in\{1,2,3,4\}$, see Fig.~\ref{fig:levelscheme}. We then choose a wavelength close to the $S_{1/2}\leftrightarrow P_{1/2}$ transition, which results for our choice of encoding $\Delta_{N,0} \gg \Delta_{N,j}$ for $j\neq 0$. This allows us to neglect differences in phase shifts between $D_{5/2}$ levels.
We then interleave $d$ applications of $U_\mathrm{LS}(t_\mathrm{g})$ with cyclic permutations of the form $X_d = \sum_{j=0}^{d-1} \ket{j+1\ \mathrm{(mod}\ d) }\bra{j}$, where the populations of each level are transferred to the level with the next higher index. This ensures that each logical state spends an equal amount of time in each physical energy level. After application of the sequence $G = (X_d U_\mathrm{LS}(t_\mathrm{g}))^d$ we find the phases $\widetilde\phi_{jk}$ imprinted on the states $\ket{jk}$
\begin{equation}
    \widetilde\phi_{jk} = \begin{cases}
    \sum_{k=0}^{d-1} \phi_{kk}\quad &\mathrm{if}\ k=j\\
    \sum_{k=0}^{d-1}\sum_{j<k} \phi_{jk}\quad &\mathrm{else} ,
\end{cases}
\label{eq:GeometricPhases}
\end{equation}
where $\phi_{jk}$ refer to the phases from the constituent $U_\mathrm{LS}(t_\mathrm{g})$ pulses. For $d=2$ the sequence $(X_2 U_\mathrm{LS}(t_\mathrm{g}))^2$ corresponds to the standard qubit light-shift gate with spin echo~\cite{Clark2021}. Equation~\eqref{eq:GeometricPhases} shows that after symmetrization we are left with only two different phases, one for the cases where both ions are in the same state and one where the ions are in different states, see Fig.~\ref{fig:gatescheme}. Hence, up to a global phase, the qudit light-shift gate operation $G(\theta)$ can be described for all $d$ by 
\begin{equation}
    G(\theta): \begin{cases}
    \ket{jj}\rightarrow \ket{jj}\quad &\\
    \ket{jk}\rightarrow \exp(i\theta)\ket{jk}  \quad &\mathrm{if}\ j\neq k .
\end{cases}
\label{eq:GateOperator}
\end{equation}
The operator $G(\theta)$ natively generates genuine qudit entanglement as opposed to merely embedding qubit-level entanglement in a larger Hilbert space~\cite{Ringbauer2021,Senko2015}. This will enable the generation of high-fidelity qudit entanglement with a single gate operation. In the experiment $\theta$ can be chosen freely by simultaneously varying the gate detuning $\delta$ and the gate time $t_g$ or changing the laser power in order to adjust the light shifts $\Delta_{N,j}$. A more detailed derivation of Eq. \eqref{eq:GateOperator} for the case of a qubit and qutrit is found in the supplementary material.

The gate is performed on two $^{40}$Ca$^+$ ions, trapped $\SI{110}{\micro m}$ above the surface of a segmented surface Paul trap. The ion trap is mounted inside a liquid Helium flow-cryostat and is operated at $\sim \SI{35}{K}$~\cite{brandl2016cryogenic}. Static potentials applied to a set of DC electrodes confine the ions along the axial direction with an axial COM mode frequency of $\omega_\mathrm{COM}/2\pi \approx \SI{1,1}{MHz}$. Radio-frequency potentials create radial confinement with frequencies $\{\omega_x, \omega_y\}/2\pi \approx \{\SI{3,5}{MHz},\SI{3,2}{MHz}\}$. Outside the vacuum chamber, a pair of Helmholtz coils aligned $\SI{45}{\degree}$ with respect to the axial trap direction generates a magnetic field of $\sim \SI{3,6}{G}$, defining the quantization axis. The $S_{1/2, m_j = -1/2}$ ground state and the sub-levels of the $D_{5/2}$ manifold are used for encoding the qudits, see Fig.~\ref{fig:levelscheme}. 
Single-qudit rotations between $\ket{0}$ and $\ket{i}$ are implemented using a $\SI{729}{nm}$ laser with a linewidth of $\sim \SI{10}{Hz}$.

The interaction of Eq.~\eqref{eq:LightShiftHamiltonian} is generated by a pair of perpendicular laser beams, one of which is parallel and the other perpendicular to the applied magnetic field, each with a waist of approximately $\SI{45}{\micro m}$, see Fig.~\ref{fig:beamgeometry}. In this configuration, the difference wavevector of the two beams is parallel to the axial trap direction to only couple to the ions' axial motion. The two beams are derived from a single frequency-doubled Titanium-Sapphire laser at a wavelength of $\lambda \approx \SI{401.2}{nm}$, approximately $\SI{8.1}{THz}$ red-detuned from the $S_{1/2}\leftrightarrow P_{1/2}$ transition. This results in low scattering errors on the order of $10^{-4}$ for an LS gate pulse with a duration of $t_\mathrm{g} \sim \SI{35}{\micro s}$. The LS force on the state $\ket{0}$ is maximized by choosing both beams to be vertically polarized. The detuning between the beams is chosen as ($\omega_\mathrm{COM} + \delta$) to couple primarily to the axial COM motion.

Maximizing the differential light shift for a given beam intensity requires the inter-ion distance to be an integer or half-integer multiple of the period of the travelling wave pattern created by the LS gate beams~\cite{Hughes2020}.
Imperfect spacing decreases the phase difference between equal and unequal states during the application of $U_\mathrm{LS}$ on the two ions, thus reducing the achievable gate speed for a fixed beam intensity. Since most error sources scale with the gate duration, correctly choosing the spacing is crucial for achieving low error rates.
Experimentally, we adjust the inter-ion spacing by varying the voltages on the trap electrodes which create the confinement in the axial direction. The ion spacing is calibrated by initialising the ions in $\ket{00}$ and applying a resonant ($\delta = 0$) LS gate pulse with variable time. If the spacing is set to a half-integer multiple of the standing wave, the breathing mode is excited to a coherent state, whereas the motion of the COM mode remains unaffected. For an integer spacing the relation between the motional modes is inverted. The motional state is read out by measuring the excitation when shelving the ions on the respective red sideband of the $S_{1/2}\leftrightarrow D_{5/2}$ transition. By observing excitation of only the breathing mode, we infer that the inter-ion distance is set appropriately, and the unwanted phase accumulation of the equal states is minimized.

After Doppler cooling, the ions' axial motional modes are cooled to around $\num{0.1}$ quanta by resolved sideband cooling. We then initialize the ions in the $\ket{00}$ state via optical pumping. 
We create an equal superposition of all qudit states by applying the operator
\begin{equation}
    P = \prod_{j=1}^{d-1} R^{0,j}(\vartheta_j,0),
    \label{eq:PreparationOperator}
\end{equation}
with rotation angle $\vartheta_j = 2\arcsin(1/\sqrt{j+1})$ and
\begin{equation}
    R^{j,k}(\vartheta, \phi) = \exp\left(-i\dfrac{\vartheta}{2}\left(\sigma^{j,k}_1(\phi)+\sigma^{j,k}_2(\phi)\right)\right)
\end{equation}
where $\sigma^{j,k}_N(\phi) = \cos(\phi)\sigma_x^{j,k}+\sin(\phi)\sigma_y^{j,k}$ denotes the rotation on ion $N$ on the transition $\ket{j}\leftrightarrow\ket{k}$ for Pauli matrices $\sigma_x$,$\sigma_y$. Each rotation $\sigma^{j,k}_N(\phi)$ is implemented by a resonant $\SI{729}{nm}$ laser pulse. Here, $\phi$ is determined by the laser phase. 
We then apply the sequence $(X_d U_\mathrm{LS}(t_\mathrm{g}))^d$, where the generalised spin echo $X_d$ is implemented by a sequence of $\pi$-pulses on the $\ket{0}\leftrightarrow\ket{j}$ transitions. For $d>2$ the phase of the first laser pulse of each $X_d$ is shifted by $\pi$ to decrease the errors due to over-rotation of the local pulses.
After the final permutation, we apply the conjugate $P^\dag$ of the initial preparation sequence. 
Up to dimension $d=4$, this leaves the system in a maximally entangled state of the form $\ket{\Psi_d} = \sum_j^{d-1} \ket{jj}/\sqrt{d}$, whereas for $d=5$ the sequence will result in the state $\ket{\Psi_5} = (3\ket{00}+2\sum_{j=1}^{4}\ket{jj})/5$. For higher-dimensional qudits, multiple applications of the gate are required to achieve maximal entanglement. 

\begin{figure}[ht!]
    \begin{center}
    \includegraphics[width=1\columnwidth]{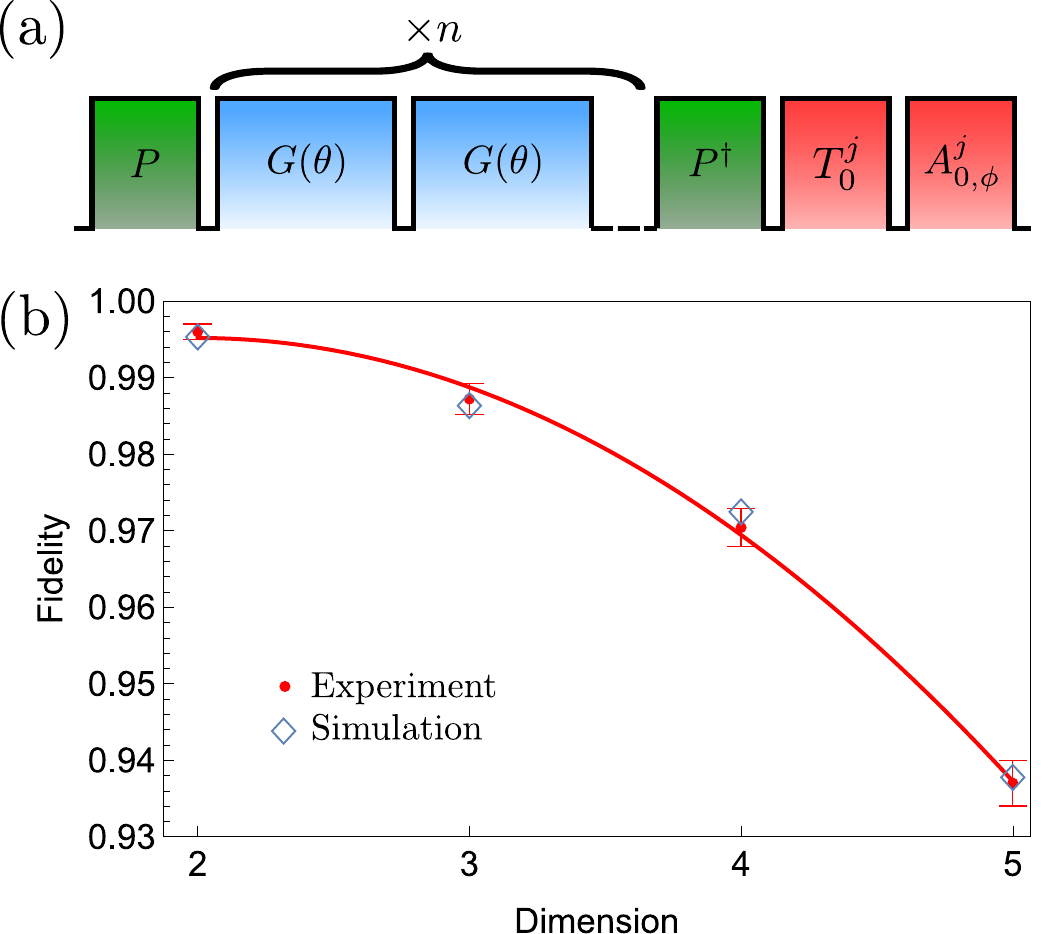}
    \caption{(a) \textbf{Pulse sequence for gate decay.} Two ions initialised in $\ket{00}$ are rotated into an equal superposition of all states by applying the operator $P$ with the $\SI{729}{nm}$ laser. 
    After applying the gate operator $G(\theta)$ a variable number of times $n$ a reversed preparation pulse $P^\dag$ is applied. The populations of the resulting state are measured by a set of transfer pulses $T_0^j$, which are resonant $\pi$ pulses between $\ket{0}\leftrightarrow\ket{j}$ to transfer the state $\ket{j}$ to the $S_{1/2}$ manifold, allowing us to distinguish the qudit states.
    An analysis pulse $A_{0,\phi}^j$ consisting of a resonant $\pi/2$ pulse between $\ket{0}\leftrightarrow\ket{j}$ with variable phase $\phi$ is used to measure the coherence between the $\ket{0}$ and $\ket{j}$ levels. Combined with the transfer pulses, all pairwise coherences can be measured.
    (b) \textbf{Qudit gate fidelity.} The average gate fidelities, shown in orange, are extracted from fits to the fidelity decay when applying multiple gates between encoding and decoding pulses. The error bars correspond to 1 standard deviation in the fit parameters. A quadratic curve has been fitted to the data to highlight the empirically observed scaling of the fidelity with dimension. The simulated fidelities from a detailed noise model are shown in blue, see supplementary information for details.
    }
    \label{fig:decays}
    \end{center}
\end{figure}

We can directly estimate the state fidelity and the amount of entanglement of the generated states from the relative amplitudes of the components $\ket{ii}$, as well as their pairwise coherences. Experimentally, the population of the $\ket{00}$ state can be measured by driving the $S_{1/2}\leftrightarrow P_{1/2}$ transition with a $\SI{397}{nm}$ laser and collecting the fluorescence photons on a photo-multiplier tube. Using additional $\pi$-pulses $T_0^j = R^{0,j}(\pi,0)$, the same procedure gives access to all components $\ket{jj}$. The coherence terms between the states $\ket{00}$ and $\ket{jj}$ are estimated by applying a $\pi/2$-pulse $A_{0,\phi}^j = R^{0,j}(\pi/2,\phi)$ with variable phase $\phi$ before performing the fluorescence readout. Applying $T_0^k$ before $A_{0,\phi}^j$ allows us to measure the coherence between the states $\ket{kk}$ and $\ket{jj}$. We then extract the coherence between the two terms from the parity oscillations by Bayesian parameter estimation, which accounts for measurement statistics and guarantees that the results stay physically possible.

The observed fidelity is affected by state-preparation-and-measurement (SPAM) errors, including the pulses $P$, $P^\dag$ the transfer pulse $T_0^j$ and analysis pulse $A_{0,\phi}^j$.
In order to separate the errors from SPAM and gate $G(\theta)$ for each dimension $d$, we insert up to $9$ applications of $G(\theta)$ between the pulses $P$ and $P^\dag$ (See Fig.~\ref{fig:decays} (a)), compute the state fidelity for each $n$ that results in an entangled state and fit an exponential decay to estimate the fidelity of a single gate. Such repeated gate applications, however, are also sensitive to the presence of non-Markovian noise in our system that leads to deviations from purely exponential decay. The extracted fidelities should thus be interpreted as an estimate for the SPAM corrected average gate performance over a sequence of length $n$.

We apply this procedure for $d=2,3,4,5$ and obtain fidelities of $99.6(1)\%, 98.7(2)\%,97.0(2)\%,93.7(3)\%$ respectively. While the intrinsic limits on gate fidelity due to finite state lifetime and Raman beam scattering depend only weakly on $d$ (see supplemental information), the measured gate performance degrades quadratically with dimension as seen in Fig.~\ref{fig:decays} (b). This can be understood if the total gate error is dominated by the errors of the local pulses, since their number increases quadratically with qudit dimension, whereas the number of entangling pulses increases linearly. Indeed this is confirmed by a numeric error model that computes the expected decay data using all independently measured error sources as inputs (see supplementary information), reproducing the observed data with very good agreement (blue points in  Fig.~\ref{fig:decays} (b)). This model suggests in particular that, for $d=2$ the gate fidelity is limited by the motional coherence time of the ion, but for $d=4,5$ the dominant error source becomes slow frequency noise that causes dephasing of the local operations. This result suggests that the gate fidelity in higher dimensions can be significantly improved if technical noise sources such as magnetic field noise contributing to the aforementioned local operation dephasing or Rabi frequency fluctuations can be suppressed. 

\begin{figure}[th]
    \begin{center}
    \includegraphics{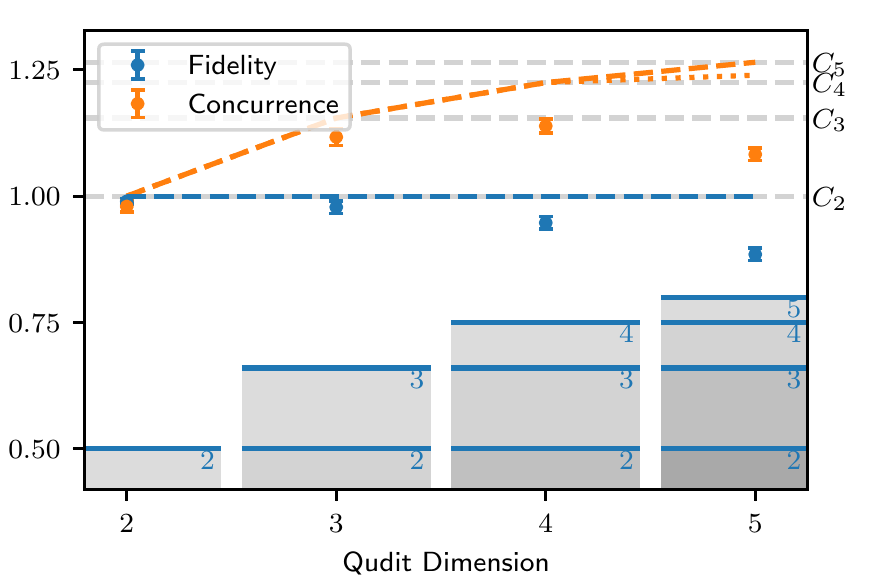}
    \caption{\textbf{Generation of genuine qudit entanglement} The measured state fidelity for $d=2,3,4,5$ is shown as blue data points and the corresponding concurrence as orange squares. The ideal values for maximally entangled states (the experimental target states) are shown as dashed (dotted) lines. The grey shaded bars represent the lower bound on the fidelity (blue data points) for certifying maximal Schmidt number entanglement, while the grey horizontal lines indicate the maximal concurrence values $C_d$ in dimension $d$. Error bars correspond to one standard deviation of experimental shot noise.}
    \label{fig:entanglement}
    \end{center}
\end{figure}

We furthermore evaluate the entanglement properties \cite{friis2019} of the states generated by a single application of the gate, including the Schmidt number, Concurrence, and Entanglement of Formation, see Fig.~\ref{fig:entanglement} and supplementary information for details. We find that the Schmidt number in each dimension is maximal, indicating the presence of genuine qudit entanglement. Crucially, the concurrence for all states with $d>2$ significantly exceeds the maximal possible value for any qubit state. Interestingly, for the ququart $(d=4)$ and ququint $(d=5)$ we do not exceed the maximum possible concurrence for qutrits $(d=3)$, highlighting that it becomes increasingly challenging to generate maximal qudit entanglement in higher dimensions.

We have demonstrated an experimental realisation of a gate that directly generates native qudit entanglement between two trapped ions. The major difference between previously demonstrated qudit entangling schemes \cite{Ringbauer2021} is that the entanglement between the multiple qudit levels is generated in a single application of the gate rather than through repeated applications of qubit entangling operations that generate only pairwise entanglement between states. This results in crucial practical benefits. In particular, the calibration of our native qudit gate compares favourably to multiple applications of a pairwise entangling M{\o}lmer-S{\o}rensen gate. In the case of the latter, the gate requires careful adjustment of the entangling laser control parameters for each of the desired $\ket{S}\leftrightarrow \ket{D}$ transitions including compensation for light shifts and induces undesired phase shifts on spectator levels, which have to be tracked. For our qudit phase gate, increasing the dimensionality of the entangling space just requires one extra local operation per additional $D_{5/2}$ sub-level to be calibrated and the power of the gate laser to be adjusted by a known analytical ratio that is not sensitive to qubit frequency shifts from light shifts or otherwise, since the gate beam is far off-resonant. Moreover, while we demonstrated a highly symmetrized version of the gate, exploiting different light shifts on different ground- and excited state levels allows for a wide range of gate actions, accessible through local operations alone.

\section{\label{sec:Ack}Acknowledgements}
This project has received funding from the European Union’s Horizon 2020 research and innovation programme under the Marie Skłodowska-Curie grant agreement No 840450. 
We gratefully acknowledge support by the EU Quantum Technology Flagship grant AQTION under Grant Agreement number 820495, and by the US Army Research Office through Grant No. W911NF-14-1-010 and W911NF-21-1-0007. We also acknowledge funding by the ERC-2021-COG 101043705.
We also acknowledge funding by the Austrian Science Fund (FWF), through the SFB BeyondC (FWF Project No. F7109), the START prize (Y879-N27), by the Austrian Research Promotion Agency (FFG) contract 872766, and by the IQI GmbH.
The research is also based upon work supported by the Office of the Director of National Intelligence (ODNI), Intelligence Advanced Research Projects Activity (IARPA), via the US Army Research Office Grant No. W911NF-16-1-0070.

\bibliographystyle{apsrev4-2}
\bibliography{biblio}
\clearpage

\onecolumngrid
\setcounter{figure}{0}
\setcounter{equation}{0}
\setcounter{table}{0}
\setcounter{section}{0}
\makeatletter 
\renewcommand{\theequation}{A\@arabic\c@equation}
\renewcommand{\thefigure}{A\@arabic\c@figure}
\renewcommand{\thetable}{A\@arabic\c@table}
\renewcommand{\thesection}{A\@Roman\c@section}

\makeatother

\begin{center}
{\bf \large Appendix \\
Native qudit entanglement in a trapped ion quantum processor}
\label{SI}
\end{center}
\medskip 

\section{\label{sec:AppendixA} Gate action in the case of the qubit and qutrit}
In this section we explicitly show how the phase acquisition in two and three dimensions leads to entanglement. For a qubit, the gate acts as follows:
\begin{align*}
   & (X_2 U_\mathrm{LS}(t_\mathrm{g}))^2\ket{00}\rightarrow \exp(\phi_{00}+\phi_{11})\ket{00} \\
   & (X_2 U_\mathrm{LS}(t_\mathrm{g}))^2\ket{01}\rightarrow \exp(\phi_{01}+\phi_{10})\ket{01} \\
   &  (X_2 U_\mathrm{LS}(t_\mathrm{g}))^2\ket{10}\rightarrow \exp(\phi_{10}+\phi_{01})\ket{10} \\
   & (X_2 U_\mathrm{LS}(t_\mathrm{g}))^2\ket{11}\rightarrow \exp(\phi_{11}+\phi_{00})\ket{11}, \\
\end{align*}
with
\begin{equation}
    \phi_{jk} = \dfrac{\pi\eta^2(\Delta_{0,j}e^{i\varphi_0}+\Delta_{1,k}e^{i\varphi_1})^2}{2\delta^2}.
\end{equation}
By choosing the ion spacing appropriately as described in the main text, the spatial phase of ion $N$ can be described as $\varphi_N = N\pi$.
With the additional assumption that the ions are equally illuminated within the travelling wave, we find $\Delta_{0,j} = \Delta_{1,j}$ for all $j$ and therefore $\phi_{00} = \phi_{11} = 0$. The asymmetric states thus acquire a phase relative to the symmetric states. An equal superposition of all four states is mapped to a maximally entangled state when the condition $\phi_{01} = \phi_{10} = \pi/4$ is fulfilled. This can, for example, be achieved by changing the coupling strength $\Delta_{n,j}$ through changing the power of the laser beams, or by changing the detuning from the motional mode $\delta_\mathrm{g}$. It should be noted that the condition for the spatial phase is not a strict requirement, as a maximally entangling gate is still possible as long as $\varphi_n \neq N\pi/2$, albeit a higher laser power is required to compensate for the phase acquisition of $\phi_{00}$ and $\phi_{11}$. 

For the case of a qutrit we find 
\begin{align*}
    &(X_3 U_\mathrm{LS}(t_\mathrm{g}))^3\ket{00}\rightarrow \exp(\phi_{22}+\phi_{11}+\phi_{00})\ket{00} \\
    &(X_3 U_\mathrm{LS}(t_\mathrm{g}))^3\ket{01}\rightarrow \exp(\phi_{20}+\phi_{12}+\phi_{01})\ket{01} \\
    &(X_3 U_\mathrm{LS}(t_\mathrm{g}))^3\ket{10}\rightarrow \exp(\phi_{02}+\phi_{21}+\phi_{10})\ket{10} \\
    &(X_3 U_\mathrm{LS}(t_\mathrm{g}))^3\ket{11}\rightarrow \exp(\phi_{00}+\phi_{22}+\phi_{11})\ket{11}\\
    &(X_3 U_\mathrm{LS}(t_\mathrm{g}))^3\ket{12}\rightarrow \exp(\phi_{01}+\phi_{20}+\phi_{12})\ket{12} \\
    &(X_3 U_\mathrm{LS}(t_\mathrm{g}))^3\ket{20}\rightarrow \exp(\phi_{12}+\phi_{01}+\phi_{20})\ket{20} \\
    &(X_3 U_\mathrm{LS}(t_\mathrm{g}))^3\ket{21}\rightarrow \exp(\phi_{10}+\phi_{02}+\phi_{21})\ket{21} \\
    &(X_3 U_\mathrm{LS}(t_\mathrm{g}))^3\ket{22}\rightarrow \exp(\phi_{11}+\phi_{00}+\phi_{22})\ket{22}.\\    
\end{align*}
If we again make the assumption that the ions are equally illuminated, we find  $\phi_{jk} = \phi_{kj}$ for all $j,k$. It then becomes evident that the states $\ket{00},\ket{11},\ket{22}$ acquire a relative phase compared to all the other states. Genuine multipartite entanglement is obtained if $\phi_{01}+\phi_{02}+\phi_{12} = 2\pi/3$. The phase acquisition during the gate is graphically illustrated in Fig.~\ref{fig:gatescheme}.

\clearpage
\section{Error Model}\label{sec:AppendixErrorModel}
The gate error analysis is based on numerically integrating the Lindblad master equation with collapse operators that describe motional heating and motional dephasing. All other noise sources (with the exception of the local gate laser) are assumed to be slow on the time scale of one experiment and thus are treated as static offsets sampled from a normal distribution with standard deviations listed in table  \ref{table:errorbudget} and averaged over 100 runs of the numerical integration. All input values are measured independently using the ions as probes with appropriate techniques (e.g. sideband thermometry, Ramsey spectroscopy, Rabi spectroscopy, Stark shift measurements etc.). When analysing the 729 laser intensity noise, we discovered further pulse area variations between subsequent $\pi$ pulses on timescales comparable to those presented during the gate operation. We thus include an additional \textit{fast} noise parameter that samples a new Rabi frequency for each subsequent local pulse during a single gate simulation run. The frequency noise for the local operations contains a contribution from both the finite 729 nm laser linewidth and magnetic field noise. Since each level in our qudit has a different magnetic field sensitivity, we scale the value in table~\ref{table:errorbudget} by empirically measured coherence times on the different transitions. 

Since the action of these slow noise components is non-Markovian we will observe a different error per gate depending on how many gates we apply in succession.
To be consistent with the decay fit method that we use to correct for SPAM and extract the fidelity from the experimental data, all simulation fidelities are extracted using an exponential fit to the same amount of gates as applied in figure \ref{fig:decays}.

The intrinsic two-qubit gate fidelity due to scattering and state decay can be calculated analytically following the analysis in \cite{Sawyer2021}. We additionally extend the $D$ state error to be determined not just by the gate length, but also the total time spent applying our generalised spin echo local operations. Extending this analysis to higher dimensions can be significantly simplified since the contributions to the scattering error arising from scattering (both elastic and inelastic) from the $D$ states are negligible. We thus obtain the $d>2$ errors by scaling the qubit error linearly by the appropriate gate time and laser intensity and correct for the fraction of the initial superposition state present in the $S$ state for $d>2$ relative to the qubit. 

\begin{table*}[h!]
\begin{centering}
\begin{tabular}{c c c | c| c | c | c } 
 \hline \hline
 Error source & Error value & Unit &   &  Infidelity &  &  \\
  &  &  & $d=2$   & $d=3$ & $d=4$ & $d=5$ \\
 [0.5ex] 
 \hline
Motional heating rate& 15 & ph/s & $1.3\times10^{-4}$& $2.2\times10^{-4}$ & $4\times10^{-4}$ &  $7\times10^{-4}$\\ 

Motional coherence& 16 & ms & $1.2\times10^{-3}$ & $2.2\times10^{-3}$ & $5\times10^{-3}$ & $5\times10^{-3}$  \\

Motional mode occupation & 0.1 & ph & $2\times10^{-4}$ & $3\times10^{-4}$ & $4\times10^{-4}$ & $1.3\times10^{-3}$\\

Gate Rabi frequency & 1 & \% & $2.2\times10^{-4}$ & $4 \times10^{-3}$ & $8\times10^{-3}$& $7\times10^{-3}$ \\

Slow local Rabi frequency & 0.6 & \% & $3\times10^{-4}$ & $4\times10^{-4}$ & $6\times10^{-4}$ & $3\times10^{-3}$ \\ 

Fast local Rabi frequency & 0.7 & \% & $2.5\times10^{-4}$ & $1\times10^{-3}$ & $1.4\times10^{-3}$& $1.6\times10^{-3}$\\ 

Local Rabi  imbalance  & 1 & \% & $4\times10^{-4}$  & $6 \times10^{-4}$ & $1.5\times10^{-3}$& $6\times10^{-3}$\\  

Gate laser frequency noise    & 2$\pi\times 200$ & rad/s & $1.3\times10^{-3}$ & $1.8 \times10^{-3}$ & $3\times10^{-3}$ & $3\times10^{-3}$\\

Local operation frequency noise & 2$\pi\times 19$ & rad/s & $1.5\times10^{-6}$ & $2.3\times10^{-3}$ & $6\times10^{-3}$  & $3.3\times10^{-2}$ \\

Elastic \& inelastic scattering &   &   & $1.6\times10^{-4}$ & $2\times10^{-4}$ & $3\times10^{-4}$  & $3\times10^{-4}$ \\
$D_{5/2} $ state decay& 1  & s  & $8\times10^{-5}$ & $2\times10^{-4}$ & $5\times10^{-4}$  & $9\times10^{-4}$ \\
\hline

Total &  & & $4.2\times10^{-3}$ & $1.3\times10^{-2}$ & $2.7\times10^{-2}$  & $6.2\times10^{-2}$ \\[1ex] 
 \hline\hline
\end{tabular}
\label{table:errorbudget}
\caption{Error sources and the corresponding simulated infidelity for the gate in qudit dimension $d=2,3,4,5$ based on measured noise parameters.}
\end{centering}
\end{table*}

\clearpage
\section{Computing qudit entanglement}\label{sec:AppendixEntanglement}
We use two figures of merit to evaluate the entangling properties of the new gate. The first is the Schmidt number, also often referred to as dimensionality of entanglement \cite{friis2019}. It quantifies the minimal local dimension needed to express the correlations present in the state, or formally
\begin{align}
    r(\rho):=\inf_{\mathcal{D}(\rho)}\max_{|\psi_i\rangle\in\mathcal{D}(\rho)}\text{rank(Tr}_B|\psi_i\rangle\langle\psi_i|\text{)}\,.
\end{align}
While computing the Schmidt number is at least as hard as deciding whether a given density matrix is separable (NP-hard), there are easily computable lower bounds. In particular, the fidelity of an experimental state $\rho_{exp}$, with an entangled target state
\begin{align}
    |\psi_T\rangle=\sum_{i=0}^{d-1}\lambda_i|ii\rangle\,,
\end{align}
is bounded by it's Schmidt rank $r$ through $F(\rho_{exp},|\psi_T\rangle\langle\psi_T|)\leq \sum_{i=0}^{r-1}\lambda_i^2$. As target state we designate the maximally entangled state in $d=2,3,4$ and the expected state in $d=5$ was chosen to be $|\psi_T\rangle=0.6\ket{00}+0.4(\ket{11}+\ket{22}+\ket{33}+\ket{44})$. We get resulting fidelities of $98.9\pm0.6\%,97.8\pm1.2\%, 94.7\pm1.2\%, 88.4\pm1.2\%$, thus proving that the Schmidt number is indeed maximal in all experimentally implemented dimensions.

The second figure of merit is a smooth quantifier that is related to the entanglement cost. The entanglement of formation is a generalisation of the entanglement entropy for pure states as the average entanglement entropy of an ensemble, minimised over all possible decompositions
\begin{align}
    E_{oF}:=\inf_{\mathcal{D}(\rho)}\sum_ip_iS(\text{(Tr}_B|\psi_i\rangle\langle\psi_i|\text{)})\,.
\end{align}
It can be lower bounded in various ways, the most easily accessible in a scalable manner is through the same elements that were needed to compute fidelity. Those can be used to compute a lower bound on the Renyi-2-entropy of entanglement \cite{huber2013}. These lower bounds, however, are known to already have some drawback when for the pure target state, the Renyi-2-entropy is already different from the von Neumann entropy (as it is in our case of dimension 5). So generally, the actual entanglement of formation will be quite seriously underestimated. Nonetheless, despite the lower bound, the non-sharpness and all experimental imperfections, the entanglement of formation still is above anything even the most perfect qubit could achieve, solidifying also the asymptotic entanglement properties of our qudit phase gate. 

\begin{table*}[h!]
\begin{centering}
\begin{tabular}{c| c| c | c | c } 
 \hline \hline

  & $d=2$   & $d=3$ & $d=4$ & $d=5$ \\
 [0.5ex] 
 \hline
Fidelity & $0.989\pm0.005$& $0.978\pm0.012$ & $0.947\pm0.012$ &  $0.884\pm0.012$\\ 
Fidelity Threshold & $0.5$ & $0.66$ & $0.75$ & $0.8$ \\
Concurrence & $0.98\pm0.01$ & $1.12\pm0.02$ & $1.14\pm0.01$ & $1.08\pm0.01$ \\
Max. Concurrence & $1$ & $1.154$ & $1.224$ & $1.264$ \\
Entanglement of Formation & $0.97\pm 0.08$ & $1.50\pm 0.12$ & $1.61\pm 0.13$ & $1.26\pm0.06$\\
Schmidt number & 2 & 3 & 4 & 5\\
 \hline\hline
\end{tabular}
\label{table:statefidelities}
\caption{Fidelities and entanglement measures for the states produced by a single entangling gate for qudit dimensions $d=2,3,4,5$. The fidelity threshold corresponds to the minimum fidelity to achieve maximal Schmidt number, while the maximum concurrence gives the value for a pure maximally entangled state of a given dimension.}
\end{centering}
\end{table*}

\end{document}